\documentclass[aps,onecolumn,amsmath,showpacs,superscriptaddress,floatfix,nofootinbib,nopreprintnumbers]{revtex4-2}

\usepackage{verbatim}
\usepackage[T1]{fontenc}
\usepackage[utf8]{inputenc}
\usepackage[american]{babel}
\usepackage{epsfig}
\usepackage{graphicx}
\usepackage{booktabs}
\usepackage{multirow}
\usepackage{dcolumn}
\usepackage{amsmath}
\usepackage{amssymb}
\usepackage{mathtools}
\usepackage{amsfonts}
\usepackage{amssymb}
\usepackage{epstopdf}
\usepackage{bm}
\usepackage{siunitx}
\usepackage{braket}
\usepackage{enumitem}
\usepackage{soul}
\usepackage[table]{xcolor}
\usepackage{color}
\usepackage{transparent}
\usepackage{pifont}
\usepackage[normalem]{ulem}

\definecolor{navyblue}{rgb}{0.0, 0.0, 0.5}
\definecolor{royalblue}{rgb}{0.25, 0.41, 0.88}
\definecolor{cadmiumgreen}{rgb}{0.0, 0.42, 0.24}
\definecolor{blue-violet}{rgb}{0.54, 0.17, 0.89}
\definecolor{darkviolet}{rgb}{0.58, 0.0, 0.83}
\definecolor{orange(colorwheel)}{rgb}{1.0, 0.5, 0.0}

\usepackage[bookmarks=true,bookmarksnumbered=true,colorlinks=true,urlcolor=royalblue, citecolor=blue,linkcolor=royalblue,breaklinks=true]{hyperref}
\hypersetup{
    colorlinks=true, 
    linkcolor=royalblue, 
    citecolor=blue}

\usepackage{booktabs}
\usepackage{multirow}
\usepackage{dcolumn}
\usepackage{colortbl}

\definecolor{magenta(process)}{rgb}{1.0, 0.0, 0.56}

\definecolor{darkspringgreen}{rgb}{0.09, 0.45, 0.27}

\definecolor{royalblue(web)}{rgb}{0.25, 0.41, 0.88}

\begin{document}

\title{Shedding light on the $\Delta m^2_{21}$ tension with supernova neutrinos}

\author{Rasmi Hajjar}
\email{rasmi.hajjar@ific.uv.es}

\affiliation{Instituto de F\'{i}sica Corpuscular (IFIC), University of Valencia-CSIC, Parc Cient\'{i}fic UV, C/ Cate\-dr\'{a}tico Jos\'{e} Beltr\'{a}n 2, E-46980 Paterna, Spain}
\affiliation{Scuola Superiore Meridionale, Largo San Marcellino 10, 80138 Napoli, Italy}

\author{Sergio Palomares-Ruiz}
\email{sergiopr@ific.uv.es}
\author{Olga Mena}
\email{omena@ific.uv.es}

\affiliation{Instituto de F\'{i}sica Corpuscular (IFIC), University of Valencia-CSIC, Parc Cient\'{i}fic UV, C/ Cate\-dr\'{a}tico Jos\'{e} Beltr\'{a}n 2, E-46980 Paterna, Spain}

\date{\today}
\preprint{}

\begin{abstract}
One long-standing tension in the determination of neutrino parameters is the mismatched value of the solar mass square difference, $\Delta m_{21}^2$, measured by different experiments: the reactor antineutrino experiment KamLAND finds a best fit larger than the one obtained with solar neutrino data. Even if the current tension is mild ($\sim 1.5\sigma$), it is timely to explore if independent measurements could help in either closing or reassessing this issue. In this regard, we explore how a future supernova burst in our galaxy could be used to determine $\Delta m_{21}^2$ at the future Hyper-Kamiokande detector, and how this could contribute to the current situation. We study Earth matter effects for different models of supernova neutrino spectra and supernova orientations. We find that, if supernova neutrino data prefers the KamLAND best fit for $\Delta m_{21}^2$, an uncertainty similar to the current KamLAND one could be achieved. On the contrary, if it prefers the solar neutrino data best fit, the current tension with KamLAND results could grow to a significance larger than $5\sigma$. Furthermore, supernova neutrinos could significantly contribute to reducing the uncertainty on $\sin^2\theta_{12}$.
\end{abstract}


\maketitle

\section{Introduction}
\label{sec:intro}

We have fully entered the high-precision era in neutrino oscillation physics. Aside from some unknowns (like the neutrino mass ordering, normal or inverted, the octant of the atmospheric mixing angle $\theta_{23}$ and the value of the CP violating phase)~\cite{deSalas:2020pgw, Esteban:2020cvm, Capozzi:2017ipn} and some experimental anomalies~\cite{Acero:2022wqg}, the overall picture of light active neutrino mixing is reasonably well understood. Yet, some small tensions among different experimental results exist~\cite{deSalas:2020pgw, Esteban:2020cvm, Capozzi:2017ipn}. 

One of these tensions, a long-standing one, lies in the determination of the solar mass square difference, $\Delta m^2_{21}$. After the latest solar neutrino analysis of Super-Kamiokande (SK) and SNO data, however, this tension has slightly decreased from the $\sim 2.2\sigma$ to the $\sim 1.5\sigma$ level~\cite{Solar2022}. The preferred value by the long-baseline reactor antineutrino experiment KamLAND~\cite{KamLAND:2013rgu} is higher, $\Delta m^2_{21} = \left(7.54^{+0.19}_{-0.18}\right) \times 10^{-5}~\mathrm{eV}^2$, than the one recently obtained from solar neutrino experiments~\cite{Solar2022}, $\Delta m^2_{21} = \left(6.10 ^{+1.04}_{-0.75}\right) \times 10^{-5}~\mathrm{eV}^2$.  Historically, the origin of this tension is twofold: the absence of a spectral up-turn at low energies in solar neutrino data, expected for larger values of $\Delta m^2_{21}$,\footnote{Indeed, the latest SK solar neutrino analysis slightly favors a low-energy up-turn, although more data would be needed to confirm it~\cite{Solar2022}.} and the observation of a day-night asymmetry in solar neutrino data which is larger than expected for the value of $\Delta m^2_{21}$ obtained by KamLAND. Note, however, that whereas KamLAND results are based on antineutrino data, solar analyses use neutrino data. Therefore, reducing the tension between these two results would also further constrain the possibility of CPT violation as a potential explanation~\cite{Ohlsson:2014cha, Barenboim:2017ewj, Barenboim:2023krl}.

In this work, we consider a different approach to determine the value of $\Delta m^2_{21}$, potentially contributing to the solution of this mild tension. We study up-going core-collapsed supernova (SN) neutrinos at the future Hyper-Kamiokande (HK) detector~\cite{Hyper-Kamiokande:2018ofw}, that is, SN neutrinos reaching this detector from below, after propagating through the Earth. Since Earth matter effects are driven by a factor proportional to $E_\nu/\Delta m^2_{21}$~\cite{Lagage:1987xu, Arafune:1987cj, Notzold:1987vc, Minakata:1987fj, Smirnov:1993ku, Dighe:1999bi, Lunardini:2000sw, Takahashi:2000it, Lunardini:2001pb, Takahashi:2001dc, Fogli:2001pm, Lunardini:2003eh, Dighe:2003jg, Dighe:2003vm, Dasgupta:2008my, Guo:2008mma, Scholberg:2009jr, Borriello:2012zc}, SN neutrinos with tens of MeV, along the tail of their spectra, would experience the presence of matter most strongly. In comparison, the regeneration of the low-energy solar $\nu_e$ flux~\cite{Spiro:1986ez, Smirnov:1986ij, LoSecco:1986tt, Bouchez:1986kb, Carlson:1986ui, Cribier:1986ak, Dar:1986pj}, caused by the Earth matter effect and that sets the day-night asymmetry, only occurs at the few percent level~\cite{Super-Kamiokande:2002ujc, Super-Kamiokande:2003yed, Super-Kamiokande:2005wtt, Super-Kamiokande:2013mie, Super-Kamiokande:2016yck, Solar2022}.

Earth matter effects on SN neutrinos have been extensively studied~\cite{Lagage:1987xu, Arafune:1987cj, Notzold:1987vc, Minakata:1987fj, Smirnov:1993ku, Dighe:1999bi, Lunardini:2000sw, Takahashi:2000it, Lunardini:2001pb, Takahashi:2001dc, Fogli:2001pm, Lindner:2002wm, Lunardini:2003eh, Dighe:2003jg, Dighe:2003vm, Akhmedov:2005yt, Dasgupta:2008my, Guo:2008mma, Scholberg:2009jr, Borriello:2012zc, Hajjar:2023knk}. In particular, the impact of the value of $\Delta m^2_{21}$ was explicitly discussed long ago~\cite{Lunardini:2000sw, Lunardini:2001pb, Takahashi:2001dc, Dighe:2003jg}. Matter effects would induce a characteristic energy-dependent modulation on the incoming SN neutrino spectra. If SN neutrinos propagate through the mantle only, there would be a well-defined frequency in the inverse-energy spectrum~\cite{Dighe:2003jg}, whereas if they also cross the Earth's core, multiple frequencies would show up~\cite{Dighe:2003vm}. These frequencies depend on the neutrino energy, on the path traversed in Earth and on the effective mass-squared difference in matter, $(\Delta m^2_{21})_\oplus$. Thus, they depend on $\Delta m^2_{21}$, the mixing angle $\theta_{12}$ and the matter potential. The identification of this oscillatory behavior would allow determining $\Delta m^2_{21}$, and the presence of matter effects is critical for this task. At low energies, around the peak of the SN neutrino spectra, statistics would be higher, but matter effects are weaker. At higher energies, along the tail of the spectra, Earth matter effects are stronger, but statistics would be lower. All in all, most of the sensitivity to Earth matter effects lies at energies of tens of MeV~\cite{Hajjar:2023knk}.

In some of the previous works on the subject, the sensitivity to $\Delta m^2_{21}$ with SN neutrinos was estimated~\cite{Takahashi:2001dc, Dighe:2003jg}. Nevertheless, either the estimates were semi-quantitative~\cite{Dighe:2003jg} or the assumed values of the solar mass square difference were relatively small~\cite{Takahashi:2001dc}, $\Delta m^2_{21} \lesssim 6 \times 10^{-5}~\mathrm{eV}^2$, compared to the current global best fit, $\Delta m^2_{21} \simeq 7.5 \times 10^{-5}~\mathrm{eV}^2$~\cite{deSalas:2020pgw, Esteban:2020cvm, Capozzi:2017ipn}, driven by the KamLAND result. For such small values of $\Delta m^2_{21}$ matter effects would be stronger than expected and the obtained results would be too optimistic. Furthermore, our knowledge of other input parameters, such as the neutrino mixing angle $\theta_{12}$ and the SN neutrino spectra, has significantly improved since then. 

Future neutrino detectors, such as the Hyper-Kamiokande (HK) facility~\cite{Hyper-Kamiokande:2018ofw}, are currently under construction. Despite the fact that the main physics goals of these upcoming experiments are the extraction of the unknown neutrino mixing parameters, a galactic SN burst could lead to hundred of thousands of neutrino events with energies below 100~MeV, depending on the precise distance to Earth~\cite{Abe:2011ts}. Therefore, in the context of future detectors such as HK, it is timely to revisit the solar--KamLAND tension on $\Delta m^2_{21}$ with updated neutrino oscillation parameters and updated SN neutrino spectra from recent numerical simulations. The structure of this work is as follows. Section~\ref{sec:mattereffect} briefly describes the Earth matter effect in the propagation of SN neutrinos. Section~\ref{sec:results} presents the study of the sensitivity to $\Delta m_{21}^2$ using galactic SN neutrino observations at HK. We perform analyses for different SN neutrino spectra and different SN orientations, exploring the impact of the detector's energy resolution and the neutrino mass ordering, and we discuss how SN neutrinos could shed some light on the $\Delta m^2_{21}$ tension. Finally, in Section~\ref{sec:conclusions}, we draw our conclusions.

\section{Earth matter effects on SuperNova neutrinos}
\label{sec:mattereffect}

After neutrinos are produced inside the dense SN as approximately mass eigenstates (in matter), they travel adiabatically and exit as mass eigenstates (in vacuum). Then, they travel towards the Earth's surface, with no flavor transitions in their cosmic path. If the SN is shadowed by the Earth, SN neutrinos would propagate through its interior for a certain distance and non-trivial matter effects could take place. Since Earth matter effects on SN neutrinos could only occur at a significant level within the $\nu_1 - \nu_2$ system, they could only be experienced by electron neutrinos ($\nu_e$) in the case of inverted neutrino mass ordering (IO) and by electron antineutrinos ($\bar{\nu}_e$) in the case of normal neutrino mass ordering (NO), as the $\nu_e$ ($\bar{\nu}_e$) spectrum for NO (IO) would be mainly composed of $\nu_3$ ($\bar{\nu}_3$)~\cite{Dighe:1999bi}. 

For neutrino energies $E_\nu \lesssim 100$~MeV, oscillatory terms driven by the atmospheric mass-square difference, $\Delta m^2_{31}$, are averaged out and the three-neutrino scenario  simplifies to a two-neutrino problem~\cite{Kuo:1986sk}. The main parameter to describe Earth matter effects on SN neutrinos is~\cite{Minakata:1987fj, Smirnov:1993ku, Ioannisian:2004jk}
\begin{equation}
	\label{eq:epsilon}
	\epsilon \equiv \frac{2 \, E_\nu \, V}{\Delta m^2_{21}} \simeq 0.12 \,  \left(\frac{E_\nu}{20~\mathrm{MeV}}\right) \, \left(\frac{Y_e \, \rho}{3~\mathrm{g/cm}^3}\right) \, \left(\frac{7.5 \times 10^{-5}~\mathrm{eV}^2}{\Delta m^2_{21}}\right)  ~,
\end{equation}
where $V = \sqrt{2} \, G_F \, N_e$ is the matter potential, being $G_F$ the Fermi constant and $N_e$ the electron number density, $Y_e$ is the electron fraction and $\rho$ is the matter density. The $\epsilon$ parameter is small ($\epsilon \lesssim 0.1$) around the peak of the SN neutrino spectra ($E_\nu \sim (10 - 20)$~MeV), growing with energy along the exponential tail of the SN neutrino spectra, where even resonant Earth matter effects could take place ($\epsilon \simeq \cos 2\theta_{12}$).

Since current and future neutrino detectors are mostly sensitive to SN $\nu_e$ and $\bar{\nu}_e$ interactions, we shall focus on these spectra in the following. The $\nu_e$ and $\bar{\nu}_e$ spectra at Earth, in terms of the spectra at production, read as~\cite{Smirnov:1993ku, Dighe:1999bi}
\begin{equation}
	\label{eq:fluxD}
	F^\mathrm{D}_{\nu_e} = p\,F^0_{\nu_e} + (1 - p) \, F^0_{\nu_x} \hspace{1cm} ; \hspace{1cm} 
	F^\mathrm{D}_{\bar\nu_e} = \overline{p} \, F^0_{\bar\nu_e} + (1 - \overline{p}) \, F^0_{\nu_x}  \hspace{1cm} ,
\end{equation}
where $p$ and $\overline{p}$ are the survival probabilities with respect to the initial spectrum. The $\nu_\mu$, $\bar{\nu}_\mu$, $\nu_\tau$ and $\bar{\nu}_\tau$ spectra are approximately equal to each other and are referred to as $\nu_x$. For the cases which could experience Earth matter effects, the probabilities for neutrinos/antineutrinos after propagating a distance $L$ in a medium of constant density (approximately, the mantle of the Earth), are given by~\cite{Smirnov:1993ku, Dighe:1999bi, Fogli:2001pm}
\begin{eqnarray}
p_\oplus^\mathrm{IO} \equiv P_\oplus(\nu_2 \rightarrow \nu_e) & \simeq & \cos^2 \theta_{13} \, \left[ \sin^2 \theta_{12} + \sin2\theta_{12}^\oplus \, \sin \left(2 \theta_{12}^\oplus - 2 \theta_{12}\right) \sin^2 \left(\pi \frac{L}{\ell_\oplus}\right) \right]	~; \nonumber \\
\overline{p}_\oplus^\mathrm{NO} \equiv P_\oplus(\bar{\nu}_1 \rightarrow \bar{\nu}_e) & \simeq & \cos^2 \theta_{13} \, \left[ \cos^2 \theta_{12} - \sin2\bar{\theta}_{12}^\oplus \, \sin \left(2 \bar{\theta}_{12}^\oplus - 2 \theta_{12}\right) \sin^2 \left(\pi \frac{L}{\bar{\ell}_\oplus}\right) \right] ~, 
\label{eq:pmatter}
\end{eqnarray}
where the oscillation length and the mixing angle (for flavor states) in matter are
\begin{equation}
\ell_\oplus = \frac{\ell_0}{\sqrt{(\cos 2\theta_{12} \mp \epsilon \, \cos^2\theta_{13})^2 + \sin^2 2\theta_{12}}} ~, 
\hspace{0.5cm} \mathrm{with} \hspace{0.5cm} \ell_0 \equiv \frac{4 \pi \, E_\nu}{\Delta m^2_{21}} ~, \hspace{1cm} \mathrm{and} \hspace{0.5cm}
\sin 2\theta_{12}^\oplus = \frac{\ell_\oplus}{\ell_0} \, \sin 2\theta_{12} ~, 
\label{eq:osclength}
\end{equation}
with $- (+)$ sign corresponding to neutrinos (antineutrinos). Note that for a two-layer density distribution, relatively simple analytical expressions can also be obtained~\cite{Dighe:2003vm}. In this work we consider the Preliminary Reference Earth Model (PREM)~\cite{Dziewonski:1981xy} for the Earth density profile and we compute the transition probabilities with the publicly available \texttt{nuSQuIDS} code~\cite{Arguelles:2021twb}. The path traversed inside the Earth by SN neutrinos is determined by the zenith angle $\theta_z$ as $L = - 2 \, R_\oplus \, \cos \theta_z\equiv - 2 \, R_\oplus \, c_z$, with $c_z \in [-1,0)$, where $R_\oplus = 6371$~km is the Earth's radius. 

The first terms within the brackets in Eq.~(\ref{eq:pmatter}) represent the vacuum contribution in each case. Thus, for constant density, the Earth matter effect on the $\nu_e$ flux can be written as
\begin{equation}
\Delta F_{\nu_e} \equiv F_{\nu_e}^\oplus -  F_{\nu_e}^\mathrm{vac} = \left(p_\oplus - p_\mathrm{vac}\right) \left(F_{\nu_e}^0 - F_{\nu_x}^0\right) = \epsilon \, \cos^4 \theta_{13} \, \sin^22\theta_{12}^\oplus \, \sin^2 \left(\pi \frac{L}{\ell_\oplus}\right) \, \left(F_{\nu_e}^0 - F_{\nu_x}^0\right)  ~.
\end{equation}
The expression for $\bar{\nu}_e$ is identical but substituting all quantities by those corresponding to $\bar{\nu}_e$. Therefore, the sensitivity to Earth matter effects depends on both the so-called regeneration factor $f_\mathrm{reg} \equiv (p_\oplus - p_\mathrm{vac})$ and the difference between $\nu_e$ (or $\bar{\nu}_e$) and $\nu_x$ spectra at production or rather, after exiting the region dominated by collective effects. Notice that these spectra are the most uncertain quantities throughout this study. Indeed, inside the progenitor star, the neutrino density is so high that neutrino self-interactions could become very strong and give rise to different flavor conversions, which can be slow~\cite{Duan:2010bg, Mirizzi:2015eza, Chakraborty:2016yeg, Horiuchi:2018ofe} or fast~\cite{Tamborra:2020cul, Richers:2022zug}. In this regard, the description above of neutrino propagation is too simplistic. Nevertheless, although significant progress has been made during the last years, our understanding about the final neutrino spectra remains incomplete and a wide range of cases is possible. For instance, these non-linear effects could give rise to spectral swaps at tens of MeV (with very little impact on the sensitivity to matter effects on the $\bar\nu_e$ spectrum) or to complete flavor equipartition (fully washing out any sensitivity to matter effects), but the outcome critically depends on the many assumptions that enter the calculations~\cite{Capozzi:2022slf}. Thus, in order to cover a range of possible variations among SN neutrino spectra at the detector, we consider results from different recent numerical simulations with matter effects induced by neutrino-nucleon interactions, as in Ref.~\cite{Hajjar:2023knk}, choosing as our benchmark case the one corresponding to a progenitor star mass of 20~$M_\odot$ (with turbulence strength parameter $\alpha_\Lambda =1.25$)~\cite{Warren:2019lgb}, which we denote as \texttt{Warren20}. The instantaneous spectra are parameterized as quasi-thermal (pinched)~\cite{Keil:2002in} and we consider their total time-integrated spectra.

\section{Determination of $\Delta m^2_{21}$}
\label{sec:results}

In the following, and as aforementioned, we consider the role of the future HK detector~\cite{Hyper-Kamiokande:2018ofw} to shed light on the current $\Delta m^2_{21}$ tension with SN neutrinos. We assume the detector to be doped with Gadolinium~\cite{Beacom:2003nk} with a concentration of 0.1\%. We briefly summarize the capabilities of HK to detect SN neutrinos. Since SN bursts last only for a few seconds, backgrounds can be reduced to negligible levels, and the full mass of each tank could be used, with a total fiducial mass of 440~kton. The main detection channel at energies of tens of MeV in water-Cherenkov detectors is inverse beta decay (IBD) ($\bar{\nu}_e + p \to e^+ + n$). Nevertheless, $\nu_e$ and $\bar\nu_e$ charged-current (CC) interactions with Oxygen are also relevant, in particular, if propagation inside the SN is adiabatic and the neutrino mass ordering is IO (such that matter effects would occur for neutrinos), and therefore we include also these channels. For the details of the calculation of event distributions at HK and how different channels are accounted for, we refer the reader to Ref.~\cite{Hajjar:2023knk}.\footnote{Note, however, that here we do not include the neutrino-electron elastic scattering contribution, since it is negligible at the energies where matter effects are most important.} We combine the different detection channels into two samples: (i) 90\% of IBD events and (ii) unidentified 10\% of IBD events plus events from $\nu_e$ and $\bar{\nu}_e$ CC interactions with Oxygen. We then perform a binned extended maximum likelihood analysis, neglecting backgrounds. We also assume that the relevant ($\bar\nu_e$ and $\nu_e$) SN neutrino spectra could be determined, either by another detector or by extrapolating to higher energies the HK measurements at energies around the peak of the spectra. Indeed, the HK detector will be able to discriminate among SN neutrino emission models with high significance~\cite{Hyper-Kamiokande:2021frf} and the high-energy tails are expected to be reasonably well described by quasi-thermal spectra~\cite{Tamborra:2012ac}.\footnote{Although neutrino shock acceleration could potentially create non-thermal high-energy tails in the $\nu_\mu$ and $\nu_\tau$ (and $\bar{\nu}_\mu$ and $\bar{\nu}_\tau$) spectra~\cite{Nagakura:2020gls}, they might be disentangled below 100~MeV if timing information is used.} We have explicitly checked that HK could distinguish among the different SN neutrino spectra we consider at high confidence, and also that allowing for smooth variations (e.g., including a tilt along the tail over the default spectrum) at a level of tens of percent has little impact on our results. Note that Earth matter effects would give rise to a characteristic oscillatory pattern on the otherwise smooth energy spectra, so in principle, even with a single detector, these effects might be observable~\cite{Dighe:2003jg, Dighe:2003vm, Akhmedov:2005yt, Borriello:2012zc}. 

We use the following total log-likelihood-ratio,
\begin{equation}
	\label{eq:chi2}
	\Delta \chi^2 \left(\Delta m^2_{21}, \sin^2\theta_{12} \, ; c_z\right) = 2 \, \sum_{i,s} \left[ N_{i, s}\left(\Delta m^2_{21}, \sin^2\theta_{12} \, ; c_z\right) - N_{i, s}^\mathrm{true}(c_z) + N_{i, s}^\mathrm{true}(c_z) \, \ln\left(\frac{N_{i, s}^\mathrm{true}( c_z)}{N_{i, s}\left(\Delta m^2_{21}, \sin^2\theta_{12} \, ; c_z\right)}\right)\right] ~,
\end{equation}
where $N_{i, s}\left(\Delta m^2_{21}, \sin^2\theta_{12} \, ; c_z\right)$ is the expected number of events in the energy bin $i$ for sample $s$, for an incoming neutrino direction determined by $c_z$, with parameters $\Delta m^2_{21}$ and $\sin^2\theta_{12}$. The true expected number of events of sample $s$ in the energy bin $i$, for SN direction determined by $c_z$, is given by $N_{i,s}^\mathrm{true}(c_z)$, and we generate ``Asimov data sets'' assuming either the KamLAND or solar best fits for $\Delta m^2_{21}$ and $\sin^2\theta_{12}$. We assume the log-likelihood-ratio, Eq.~(\ref{eq:chi2}), to follow a $\chi^2$ distribution with two degrees of freedom.

Since the other neutrino oscillation parameters affect the results at a sub-leading level, the inclusion of priors on those parameters has a negligible impact. We do not include priors on $\Delta m^2_{21}$ and $\sin^2\theta_{12}$ from other measurements either. In this way, we can estimate the full power of using SN neutrinos only and discuss whether SN neutrinos could shed some light on the long-standing tension between solar and reactor data in the determination of $\Delta m^2_{21}$. 

\begin{figure}[t]
\begin{center}
\includegraphics[width=0.97\columnwidth]{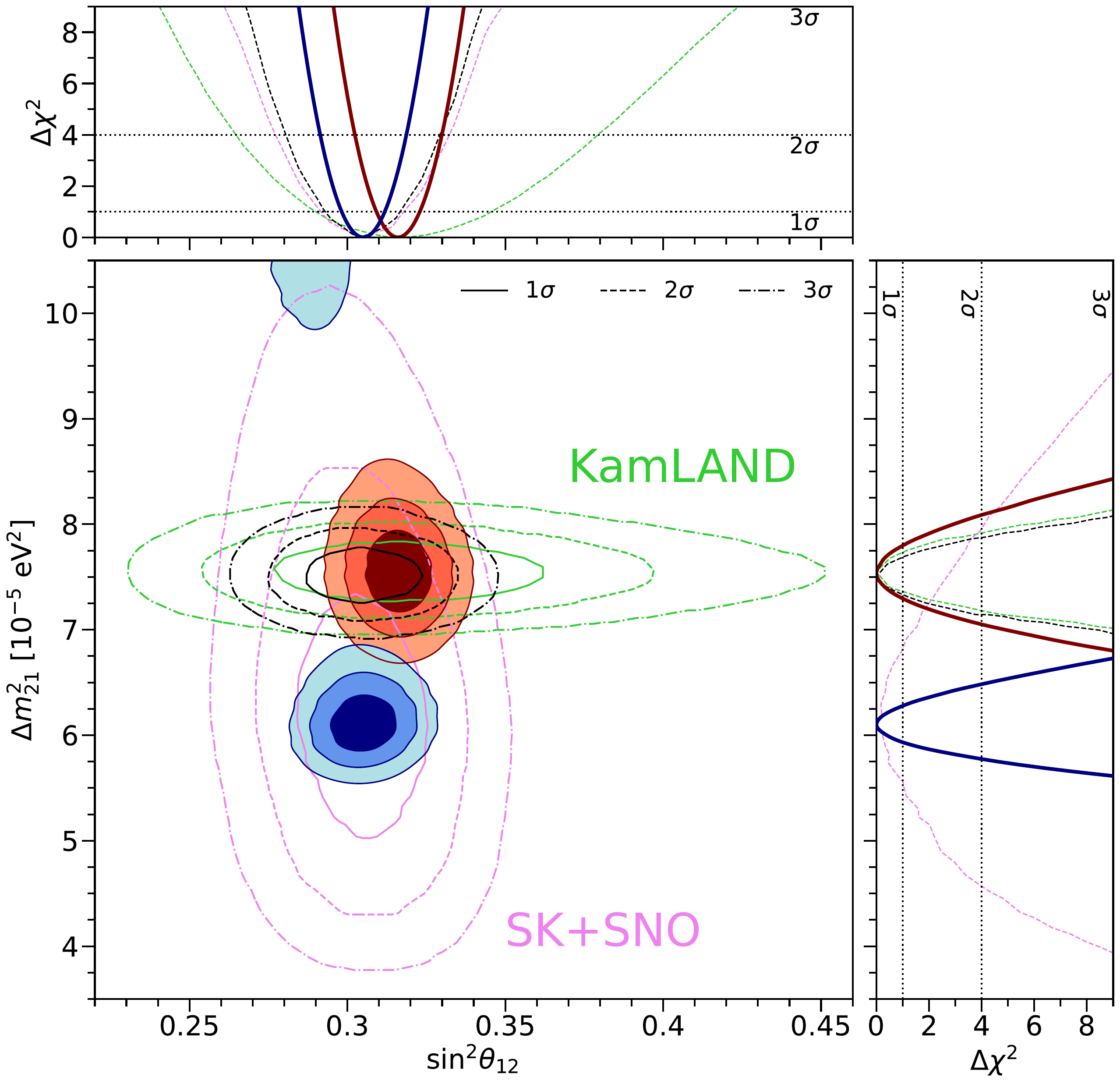} 
\caption{\textit{\textbf{Sensitivity to \boldsymbol{$\Delta m_{21}^2$} and \boldsymbol{$\sin^2 \theta_{12}$} via Earth matter effects with SN neutrinos.}} Two different sets of true values are considered: KamLAND best fit, $\Delta m^2_{21}|_{\mathrm{KL}} = 7.54 \times 10^{-5}~\mathrm{eV}^2$ and $\sin^2\theta_{12}|_{\mathrm{KL}} = 0.316$ (red hues); solar best fit, $\Delta m^2_{21}|_{\mathrm{solar}} = 6.1 \times 10^{-5}~\mathrm{eV}^2$ and $\sin^2\theta_{12}|_{\mathrm{solar}} = 0.305$ (blue hues). Regions at $1\sigma$, $2\sigma$ and $3\sigma$ CL (from darker to lighter colors) are shown. Results are obtained for NO using the \texttt{Warren20} SN neutrino spectra~\cite{Warren:2019lgb}, assuming a SN-Earth distance of 10~kpc and the SN burst to occur on the opposite side of the detector ($c_z = -1$). Also shown are the most favored regions using data from KamLAND~\cite{KamLAND:2013rgu} (green contours), solar neutrinos~\cite{Solar2022} (magenta contours) and their combination~\cite{Solar2022} (black contours).
}
\label{fig:w20_contours}
\end{center}
\end{figure}

Figure~\ref{fig:w20_contours} illustrates the sensitivity of HK to $\Delta m^2_{21}$ and $\sin^2\theta_{12}$, assuming a SN-Earth distance of 10~kpc (which is the benchmark SN-Earth distance used in this work) and also assuming that SN neutrinos cross the entire diameter of the Earth ($c_z = -1$). We consider the case of adiabatic propagation inside the SN and NO of neutrino masses (i.e., IBD is the main detection channel where matter effects could show up). We show the $1\sigma$, $2\sigma$ and $3\sigma$ confidence level (CL) regions (from darker to lighter colors) for the \texttt{Warren20} SN neutrino spectra. The results are obtained assuming two different true values for the solar neutrino mixing parameters: (i) in red filled contours, those resulting from the current KamLAND-only best fit~\cite{KamLAND:2013rgu}, $\Delta m^2_{21}|_{\mathrm{KL}} = 7.54\times 10^{-5}$ and $\sin^2\theta_{12}|_{\mathrm{KL}} = 0.316$ and (ii) in blue filled contours, those arising with the current solar data (SK+SNO) best fit~\cite{Solar2022}, $\Delta m^2_{21}|_{\mathrm{solar}} = 6.10\times 10^{-5}$ and $\sin^2\theta_{12}|_{\mathrm{solar}} = 0.305$. We also include the corresponding contours using KamLAND-only data (green contours), SK+SNO data (magenta contours) and their combination (black contours)~\cite{Solar2022}.

\begin{figure}[t]
\begin{center}
\includegraphics[width=0.97\columnwidth]{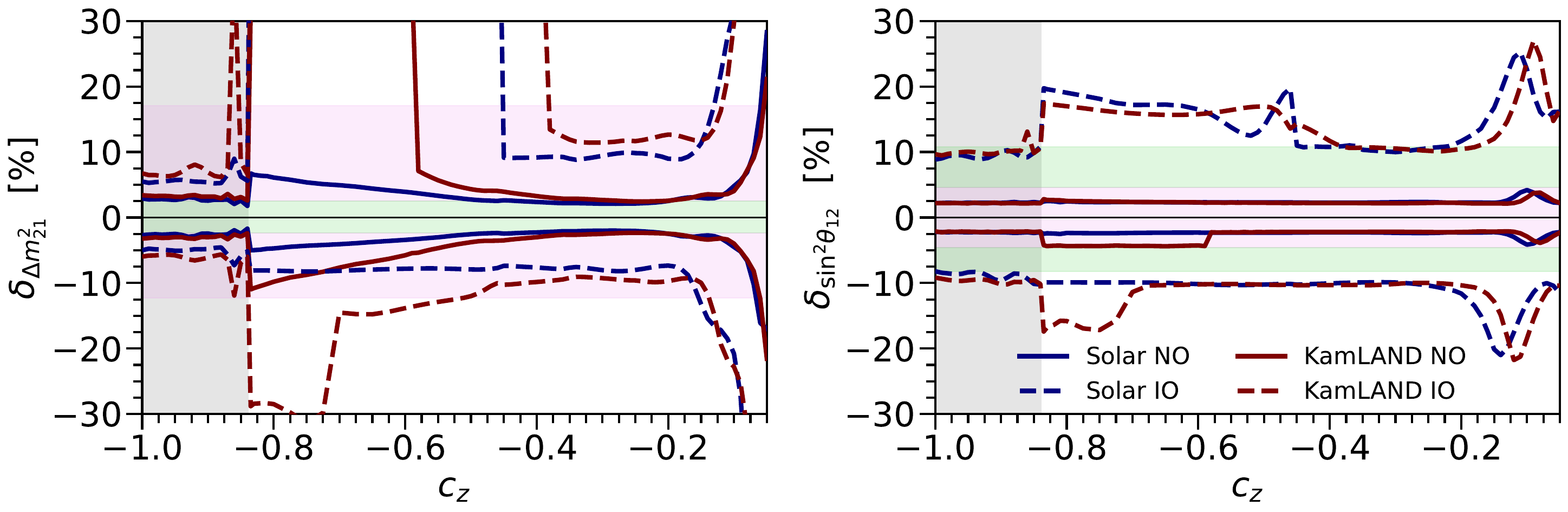} 
\caption{\textit{\textbf{Sensitivity to solar mixing parameters via Earth matter effects with SN neutrinos, as a function of the SN-detector direction, $\boldsymbol{c_z}$.}} \textit{Left panel:} $1\sigma$ uncertainty on $\Delta m^2_{21}$, $\delta_{\Delta m^2_{21}}$. \textit{Right panel:} $1\sigma$ uncertainty on $\sin^2\theta_{12}$, $\delta_{\sin^2\theta_{12}}$. Two different sets of true values are considered: KamLAND best fit, $\Delta m^2_{21}|_{\mathrm{KL}} = 7.54 \times 10^{-5}~\mathrm{eV}^2$ and $\sin^2\theta_{12}|_{\mathrm{KL}} = 0.316$ (red lines); solar best fit, $\Delta m^2_{21}|_{\mathrm{solar}} = 6.1 \times 10^{-5}~\mathrm{eV}^2$ and $\sin^2\theta_{12}|_{\mathrm{solar}} = 0.305$ (blue lines). Results are obtained for NO (solid lines) and IO (dashed lines) using the \texttt{Warren20} SN neutrino spectra~\cite{Warren:2019lgb} and assuming a SN-Earth distance of 10~kpc. Also shown are the $1\sigma$ errors using data from KamLAND~\cite{KamLAND:2013rgu} (green region) and solar neutrinos~\cite{Solar2022} (magenta region).
}
\label{fig:w20_1sigma}
\end{center}
\end{figure}

If the true best fit of $\Delta m^2_{21}$ is close to the KamLAND-driven result, SN neutrinos could constrain $\Delta m_{21}^2$ at a similar level to KamLAND data. Since the extent of the current (solar--KamLAND) tension is mainly driven by uncertainties on the solar neutrino analyses, SN neutrino data would confirm the KamLAND result, but it would barely affect the current discrepancy with solar data. Note, however, that the sensitivity to $\Delta m_{21}^2$ of SN neutrinos is based on the very same Earth matter effect as in the case of solar neutrinos, and there would be then an additional tension between SN and solar neutrino data, at a similar level to the current solar--KamLAND one. 

On the other hand, if the true value of $\Delta m^2_{21}$ is close to the solar data best fit, matter effects on SN neutrinos crossing the Earth would be stronger, as their importance scales with $E_\nu/\Delta m_{21}^2$. As a consequence, the uncertainty on $\Delta m^2_{21}$ could be significantly reduced with respect to the solar neutrino data analysis, ending up comparable to the current KamLAND uncertainty. This would imply a much bigger tension, in that case between SN antineutrinos and KamLAND reactor antineutrinos, at $\gtrsim 5\sigma$~CL.

It is important to stress that in both cases (i.e., regardless the best-fit value of $\Delta m^2_{21}$ lies closer to the KamLAND or the solar one) SN neutrino data could allow to reduce the current uncertainty on $\sin^2\theta_{12}$. Such an improvement ranges from a factor of two with respect to solar neutrino data to a factor of about four with respect to KamLAND data. Profiling over the other parameter (either $\Delta m^{2}_{21}$ or $\sin^2\theta_{12}$), the $1\sigma$~CL uncertainties on the solar neutrino mixing parameters, for these particular SN neutrino spectra and neutrino mass ordering (for $c_z = -1$), are:  
\begin{eqnarray}
\label{eq:1sig_errors}
\textrm{KamLAND best fit as true value:} \hspace{7mm} 	
\Delta m^{2}_{21} & = & \left(7.54^{+0.26}_{-0.24}\right)\times 10^{-5}~\mathrm{eV}^2 \hspace{7mm} ; \hspace{7mm}
\sin^2\theta_{12} = 0.316\pm 0.007  ~; \nonumber \\[1ex]
\textrm{Solar best fit as true value:} \hspace{7mm} 
\Delta m^{2}_{21} & = & \left(6.10^{+0.18}_{-0.17}\right)\times 10^{-5}~\mathrm{eV}^2 \hspace{7mm} ; \hspace{7mm} 
\sin^2\theta_{12} = 0.305\pm 0.007 ~.
\end{eqnarray}

The results are shown in Fig.~\ref{fig:w20_contours}. Notice,  however, that we have assumed a particular set of SN neutrino spectra, SN neutrinos to cross the entire Earth, and NO for the neutrino mass ordering. This is an optimistic scenario. Therefore, in the following we explore how results change for different trajectories through the Earth, for both neutrino mass orderings. Following Ref.~\cite{Hajjar:2023knk}, in addition to \texttt{Warren20}, we also consider other three SN neutrino spectra, \texttt{Warren9} and \texttt{Warren120}~\cite{Warren:2019lgb}, and \texttt{Garching19}~\cite{Bollig:2020phc} (see also, e.g., Refs.~\cite{Vartanyan:2019ssu, Burrows:2019zce, Nagakura:2020qhb, Wang:2022dva, Tsang:2022imn}). Indeed, among these four sets of SN neutrino spectra, the most optimistic results for $c_z = -1$ are obtained for \texttt{Warren20}.

In Fig.~\ref{fig:w20_1sigma} we show the projected $1\sigma$ uncertainties on $\Delta m^2_{21}$ and $\sin^2\theta_{12}$ (profiling over the other parameter in each case), $\delta_{\Delta m^2_{21}} \equiv \sigma_{\Delta m^2_{21}}/\Delta m^2_{21}|_\mathrm{bf}$ and $\delta_{\sin^2\theta_{12}} \equiv \sigma_{\sin^2\theta_{12}}/\sin^2\theta_{12}|_\mathrm{bf}$, under different assumptions. For comparison with Fig.~\ref{fig:w20_contours} we show the results for the \texttt{Warren20} SN spectra. Note that to obtain these results we do not scan beyond the parameter space shown in Fig.~\ref{fig:w20_contours}. In both panels of Fig.~\ref{fig:w20_1sigma}, we consider the KamLAND (red lines) and solar (blue lines) best fits as true values and depict the results for NO (solid lines) and IO (dashed lines). As expected, the best results, in particular for $\Delta m^2_{21},$ are obtained for NO and assuming the solar best fit as the true value. In this case, the expected uncertainties for most directions would be very similar to those illustrated in Fig.~\ref{fig:w20_contours} for $c_z = -1$. For the other three cases, sharp jumps on the relative errors as a function of $c_z$ are visible, which are more pronounced for $\delta_{\Delta m^2_{21}}$. They are explained by the appearance of degenerate disjoint $1\sigma$ regions in the $\sin^2 \theta_{12} - \Delta m^2_{21}$ plane, which eventually merge for longer trajectories through the mantle and shrink again if neutrinos cross the core.

On another hand, assuming the profiled log-likelihoods with KamLAND and SN data follow normal distributions (which is a reasonable approximation, see Fig.~\ref{fig:w20_contours}), the difference of $\Delta m_{21}^2$ best fits would also be normal distributed. Thus, one could test the null hypothesis of equal $\Delta m_{21}^2$, or in other words, the tension between two different measured values of $\Delta m_{21}^2$, using the test statistic:
\begin{equation}
\label{eq:z_tension}
\mu_{21} = \frac{\Delta m_{21}^2|_\mathrm{KL} - \Delta m_{21}^2|_\mathrm{SN}}{\sqrt{\sigma_\mathrm{KL}^2 + \sigma_\mathrm{SN}^2(c_z)}} ~,
\end{equation}
which is normally-distributed with mean zero and variance one. While have only explicitly specified the dependence of the variance $\sigma_\mathrm{SN}^2$ on the neutrino trajectory through the Earth ($c_z$), notice that it also depends on the SN neutrino spectra and on the neutrino mass ordering. Here, we study the case in which the value of $\Delta m_{21}^2$ measured with SN neutrinos corresponds to the current solar best fit ($\Delta m_{21}^2|_\mathrm{SN} = \Delta m_{21}^2|_\mathrm{solar}$).\footnote{As already mentioned, if the true value is closer to the KamLAND result, the tension with solar data would be dominated by uncertainties in the solar data analyses, even with future solar data at HK~\cite{Yano:2021usb}. Hence, it would be similar to the tension between solar and KamLAND data, so we do not discuss further this possibility. In that case, the interest would be knowing how much SN neutrino data could contribute to improve current KamLAND results, mainly on $\sin^2 \theta_{12}$. Of course, the constraining power of $\Delta m_{21}^2$ and $\sin^2 \theta_{12}$ would be different for different SN neutrino spectra, SN neutrino trajectories through the Earth and neutrino mass orderings, and the scenario shown in Fig.~\ref{fig:w20_contours} and quantified in Eq.~(\ref{eq:1sig_errors}) is an optimistic one.}  With the current KamLAND and solar neutrino data (i.e., $\sigma_\mathrm{SN} \to \sigma_\mathrm{solar} = 1.04 \times 10^{-5}~\mathrm{eV}^2$ in Eq.(\ref{eq:z_tension})), $(\mu_{21})_\mathrm{KL-solar} \simeq 1.4$. We also illustrate the tension using future projected uncertainties with the JUNO detector, using $\sigma_\mathrm{JUNO} = 0.024 \times 10^{-5}~\mathrm{eV}^2$~\cite{JUNO:2022mxj} instead of the KamLAND standard deviation of $\Delta m_{21}^2$, $\sigma_\mathrm{KL} = 0.18 \times 10^{-5}~\mathrm{eV}^2$. Furthermore, note that the uncertainty could also be reduced with reactor antineutrinos at HK (and even at SK doped with Gd) slightly below the KamLAND value~\cite{Choubey:2004bf, deGouvea:2020vww}.

\begin{figure}[t]
\begin{center}
\includegraphics[width=\columnwidth]{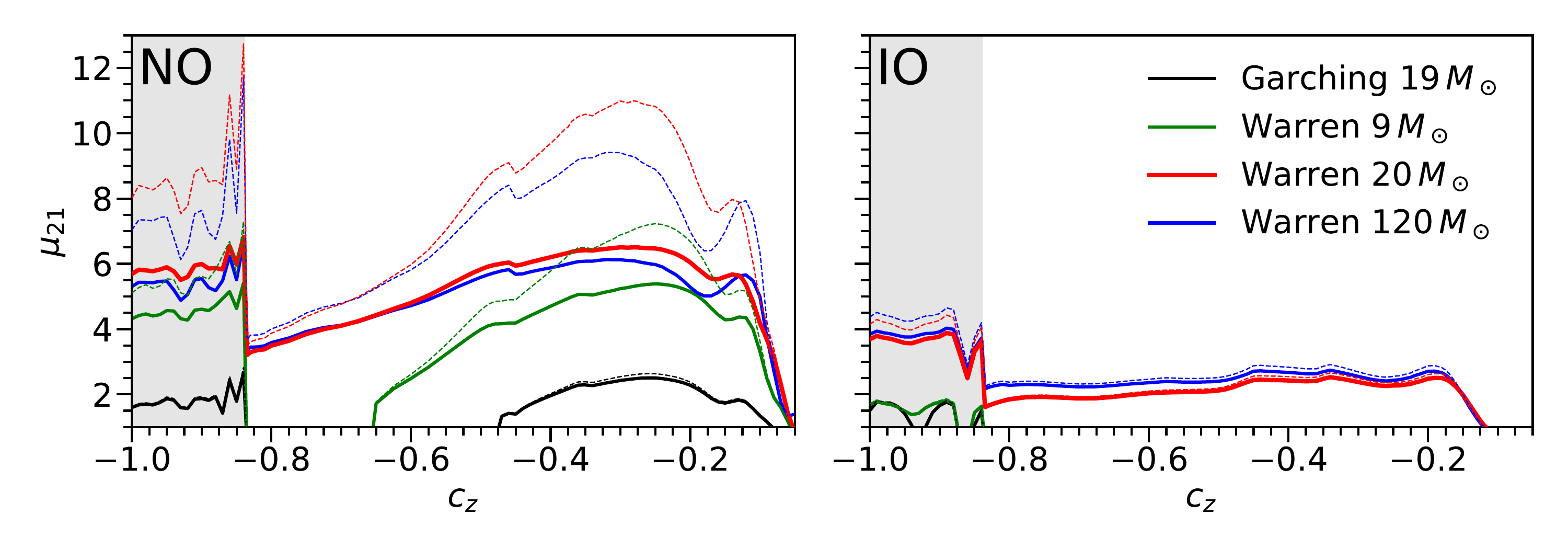} 
\caption{\textit{\textbf{Potential tension between KamLAND and future SN neutrino data at HK: dependence on the SN neutrino spectra.}} We assume the measured value is the current solar best fit, $\Delta m_{21}^2 = 6.1 \times 10^{-5}~\mathrm{eV}^2$ and show  the test statistic $\mu_{21}$ (which measures the tension in number of standard deviations), Eq.~(\ref{eq:z_tension}), as a function of the SN neutrino trajectory ($c_z$). Results for NO (\textit{left panel}) and IO (\textit{right panel}) are depicted for various SN neutrino spectra: \texttt{Warren9} (green lines), \texttt{Warren20} (red lines), \texttt{Warren120} (blue lines)~\cite{Warren:2019lgb} and \texttt{Garching19} (black lines)~\cite{Bollig:2020phc}. The tension using the projected JUNO uncertainties~\cite{JUNO:2022mxj}, instead of KamLAND's, is also depicted (dashed lines). The shaded region indicates core-crossing trajectories. We assume a SN-Earth distance of 10~kpc.}
\label{fig:tension_cz_models}
\end{center}
\end{figure}

\begin{figure}[t]
\begin{center}
\includegraphics[width=\columnwidth]{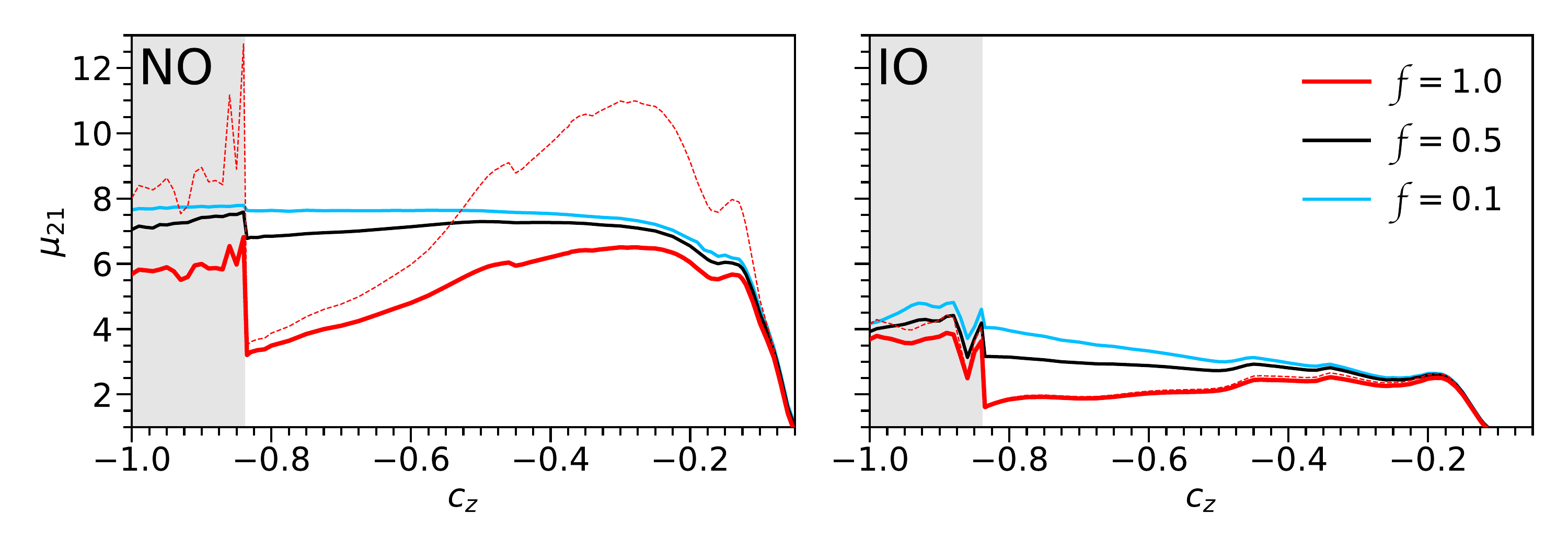} 
\caption{\textit{\textbf{Potential tension between KamLAND and future SN neutrino data at HK: dependence on the energy resolution.}} Same as Fig.~\ref{fig:tension_cz_models}, but only for the \texttt{Warren20} model, for different electron/positron energy resolutions: $\sigma_\mathrm{det} = f \, \sigma_\mathrm{HK}$, where $f =$ 0.1 (blue lines), 0.5 (black lines) and 1 (red lines). Results are shown for NO (\textit{left panel}) and IO (\textit{right panel}). The case $f = 0.1$ approximately represents the limit of perfect resolution.}
\label{fig:tension_cz_kres}
\end{center}
\end{figure}

The dependence of the test statistic $\mu_{21}$ on the neutrino trajectory is shown in Figs.~\ref{fig:tension_cz_models} and~\ref{fig:tension_cz_kres}, for NO (left panels) and IO (right panels). Whereas for NO matter effects would occur for $\bar{\nu}_e$, which give rise to the dominant event rate via IBD, for IO matter effects would take place for neutrinos, whose contribution via $\nu_e-$Oxygen interactions is not only subdominant, but it is also assumed to be indistinguishable from a fraction of the IBD events (10\% in this work) and from events induced by $\bar{\nu}_e-$Oxygen interactions. Thus, the sensitivity to matter effects using SN neutrinos at HK would be significantly better for the NO case, even if matter effects are non-resonant for antineutrinos~\cite{Hajjar:2023knk}.

In all cases illustrated in Figs.~\ref{fig:tension_cz_models} and~\ref{fig:tension_cz_kres}, a local maximum in the test statistic $\mu_{21}$ occurs for trajectories with $c_z \sim (-0.10, -0.15)$, which correspond to path lengths $\sim \ell_\oplus/2$~\cite{Bakhti:2020tcj} for neutrino energies $E_\nu \sim (50-70)$~MeV (where matter effects are very relevant). For perfect energy resolution, and mantle-crossing trajectories, the tension would grow with the path length. However, for finite energy resolution, the sensitivity to two different values of $\Delta m_{21}^2$ is reduced for longer path lengths within the mantle. This also depends on which channels are most sensitive to matter effects, which in turn depends on the neutrino mass ordering. In Fig.~\ref{fig:tension_cz_models}, for the default energy resolution and NO, $\mu_{21}$ reaches a maximum and then decreases until neutrino trajectories cross the Earth's core. Then, for those neutrinos propagating through the core, the average density is larger, so matter effects are more important at lower energies, and $\mu_{21}$ presents a jump. For IO, the sensitivity for SN neutrinos crossing the mantle has a mild dependence on the trajectory, but it is better when the core is traversed. These generic features can be seen for the four SN neutrino spectra in Fig.~\ref{fig:tension_cz_models}. The most optimistic results are obtained for \texttt{Warren20} and \texttt{Warren120}, for which $\bar{\nu}_e$ (or $\nu_e$) and $\nu_x$ spectra are quite different among each other in the relevant energy range (see Fig.~1 in Ref.~\cite{Hajjar:2023knk}). Even if the results for different models do have important differences, note that in all cases some trajectories could contribute in a significant manner to the current tension. This is also true even for the \texttt{Garching19} spectra, with very limited sensitivity to matter effects~\cite{Hajjar:2023knk}.

Figure~\ref{fig:tension_cz_models} also illustrates the effect of reducing the uncertainty on $\Delta m_{21}^2$ with reactor antineutrinos (dotted lines), that, is, reducing the value of $\sigma_\mathrm{KL}$. If it decreases by a factor of $\sim 7-8$, as it would be the case of the future JUNO detector~\cite{JUNO:2022mxj}, the denominator in Eq.~(\ref{eq:z_tension}) would have a stronger dependence on the trajectory. If $\sigma_\mathrm{SN} \lesssim \sigma_\mathrm{KL}$, then reducing the error with reactor antineutrinos (i.e.,  $\sigma_\mathrm{KL} \to \sigma_\mathrm{JUNO}$) would significantly increase the tension between the measured values of $\Delta m_{21}^2$. On the other hand, if $\sigma_\mathrm{SN} > \sigma_\mathrm{KL}$, the effect of reducing the error with reactor antineutrinos would be very mild. Thus, only for the cases for which $\mu_{21} \gtrsim 4$ with the current KamLAND uncertainty (solid lines), a significant enhancement of the tension with JUNO could be obtained.

The effect of the energy resolution is studied in Fig.~\ref{fig:tension_cz_kres}, where we consider cases with improved capabilities, parameterized as $\sigma_\mathrm{det} = f \, \sigma_\mathrm{HK}$, where $\sigma_\mathrm{HK}$ is the default electron/positron energy resolution assumed for HK. It is important to stress that this corresponds to the reconstruction of the energy of the produced positron (in $\bar{\nu}_e-$induced events) or electron (in $\nu_e-$induced events). For IBD, the width of the distribution of the final positron is ${\cal O}(2 E_\nu/m_p)$, where $m_p$ is the proton mass. For interactions with Oxygen at tens of MeV, transitions to multiple excited states would result in even larger widths~\cite{Nakazato:2018xkv}. Since the nuclear recoil energy is not measured, this actually represents a limit for the reconstruction of the neutrino energy. Indeed, for neutrino energies of tens of MeV, this width could be comparable to the electron/positron energy resolution of the detector. This implies that beyond some energy resolution, no further improvement would modify these results. This can be seen from the $f=0.5$ and $f = 0.1$ cases in Fig.~\ref{fig:tension_cz_kres}, which render very similar results. Overall, the effects of energy resolution are relatively more important for trajectories within the mantle, for path lengths $\sim 2000 - 5000$~km. 

In the case of NO, matter effects would occur for antineutrinos and hence, they would not present a resonant behavior and would be more important for higher energies than in the case of neutrinos. This results in an approximately constant value of $\mu_{21}$ for $c_z \lesssim -0.2$ if (almost) perfect electron/positron energy resolution is achieved. The worse the energy resolution, the sharper the jump at the core-mantle boundary and the more suppressed fast oscillations for long path lengths, due to energy smearing. This results in the maximum of $\mu_{21}$ around $c_z \sim -0.3$ for the default HK energy resolution assumed in this work. 

In the case of IO, matter effects would occur for neutrinos, so they could be resonant. If the energy resolution is good enough to resolve oscillations, sensitivity to matter effects would grow for longer trajectories, for which the average density is larger and hence, resonant matter effects would occur at lower energies, where statistics are higher. All in all, neutrino interactions are subdominant and unidentified IBD events represent an important background. Therefore, for IO, if adiabatic propagation inside the SN takes place, the HK sensitivity to Earth matter effects is worse than for NO. Thus, for IO the uncertainties on $\Delta m_{21}^2$ from SN neutrino data would be larger than for NO. Therefore, the HK contribution (with SN neutrinos) to solve the solar--KamLAND tension would be less significant.

\section{Conclusions}
\label{sec:conclusions}

One of the long-standing empirical tensions in neutrino oscillation physics lies in the determination of the solar mass square difference, $\Delta m^2_{21}$, by reactor antineutrino and solar neutrino measurements. Whereas the KamLAND long-baseline reactor experiment data analysis results in a value of $\Delta m^2_{21} = \left(7.54^{+0.19}_{-0.18}\right) \times 10^{-5}~\mathrm{eV}^2$~\cite{KamLAND:2013rgu}, solar neutrino data prefers a smaller value, $\Delta m^2_{21} = \left(6.10 ^{+1.04}_{-0.75}\right) \times 10^{-5}~\mathrm{eV}^2$~\cite{Solar2022}. Notice that this mild $\sim 1.5\sigma$ tension is mainly caused by the uncertainty from solar data.

In this work, we have considered the possibility to extract independently the value of $\Delta m^2_{21}$ by means of a different approach. We have studied the sensitivity of the future HK detector to $\Delta m_{21}^2-$driven matter effects on SN neutrinos propagating through the Earth. Note that the matter effect we study here is the same that sets the day-night asymmetry in solar neutrinos. There are two main differences, though. In the case of SN neutrinos, we consider higher energies (several tens of MeV) than those of solar neutrinos. In this way, Earth matter effects could be maximal for SN neutrinos, albeit along the tail of their spectra. On another hand, whereas solar data relies on the neutrino channel, in the case of SN neutrinos, both the neutrino and antineutrino channels are available, and therefore matter effects could take place for both neutrinos or antineutrinos, depending on the neutrino mass ordering.

For a galactic SN burst, were neutrino propagation inside the star adiabatic, the projected results for the HK sensitivity to $\Delta m_{21}^2$ would be more optimistic for NO. In that case, Earth matter effects would occur for antineutrinos, which are detected via IBD, the main channel at HK (at tens of MeV). In the case of IO, matter effects would occur for neutrinos, which are detected via subdominant channels at those energies, neutrino interactions off Oxygen nuclei. 

The results also depend on the SN neutrino spectra at Earth (more concretely, on the difference between the electron-flavor and the muon/tau-flavor neutrino spectra at production), on the direction of the SN with respect to the position of the detector on Earth and on the neutrino mass ordering. For a SN-Earth distance of 10~kpc, we have studied an optimistic case: the \texttt{Warren20} SN neutrino spectra, SN neutrinos crossing the entire Earth ($c_z = -1$) and NO (Fig.~\ref{fig:w20_contours}). The sensitivity to $\Delta m_{21}^2$ and $\theta_{12}$ is illustrated for both the best-fit values of $\Delta m_{21}^2$ obtained with KamLAND and solar data. If the value measured by SN neutrinos is that determined with KamLAND data, $\Delta m_{21}^2$ could be constrained at a similar level to the current KamLAND uncertainty. In this case, SN neutrinos would confirm KamLAND results, but would barely modify the solar--KamLAND tension. SN neutrinos would improve on the determination of $\sin^2 \theta_{12}$, though. If the value measured by SN neutrinos is that determined with solar neutrino data, the uncertainty could be much smaller than the current (or even future~\cite{Yano:2021usb}) solar uncertainty, since matter effects are sensitive to $E_\nu/\Delta m_{21}^2$. In this case, the tension between SN and KamLAND data could be much more worrisome than the current one between solar and KamLAND data, reaching a significance of $\sim 6\sigma$. This tension could become even larger if JUNO measurements remain compatible with the KamLAND result. 

To establish how the current solar--KamLAND tension could increase if the value measured with SN neutrinos is that determined by solar data and reactor antineutrino experiments keep on obtaining the same best fit, we have also considered different SN neutrino models and have studied the dependence on the trajectory through Earth, for both neutrino mass orderings (Fig.~\ref{fig:tension_cz_models}). The effect of the energy resolution to resolve Earth matter effects is also studied (Fig.~\ref{fig:tension_cz_kres}). In all these cases, we have also indicated how these results would change if future JUNO data remain compatible with KamLAND, but with smaller uncertainties. 

To conclude, SN neutrinos could provide an independent determination of the neutrino oscillation parameters $\Delta m_{21}^2$ and $\sin^2\theta_{12}$. Even if the results show a non-trivial dependence on the neutrino trajectory and on the precise SN neutrino spectra, we have shown that a future galactic SN explosion has the potential of rendering a competitive measurement of $\Delta m_{21}^2$, contributing to the solution of the long-standing, although currently mild, tension between solar neutrino and reactor antineutrino data, and to significantly reduce the uncertainty on $\sin^2\theta_{12}$.

\begin{acknowledgments}

This work has been supported by the Spanish MCIN/AEI/10.13039/501100011033 grants PID2020-113644GB-I00 (RH and OM) and PID2020-113334GB-I00 (SPR) and by the European Union’s Horizon 2020 research and innovation programme under the Marie Skłodowska-Curie grants H2020-MSCA-ITN-2019/860881-HIDDeN and HORIZON-MSCA-2021-SE-01/101086085-ASYMMETRY. RH is supported by the Spanish grant FPU19/03348 of MU and acknowledges partial support by the research grant number 2017W4HA7S “NAT-NET: Neutrino and Astroparticle Theory Network” under the program PRIN 2017 funded by the Italian Ministero dell’Università e della Ricerca (MUR).
The authors also acknowledge support from the Generalitat Valenciana grants PROMETEO/2019/083 and CIPROM/2022/69 (RH and OM) and CIPROM/2022/36 (SPR). SPR is also partially supported by the Portuguese FCT (CERN/FIS-PAR/0019/2021). 
\end{acknowledgments}

\bibliography{biblio}
\end{document}